\documentclass[prl,10pt,superscriptaddress,notitlepage,twocolumn,amsmath,amssymb]{revtex4-2}

\usepackage{bm}
\usepackage{braket}
\usepackage{upgreek}
\usepackage{amsmath}
\usepackage{graphicx}
\usepackage{booktabs}
\usepackage{siunitx}
\usepackage{float}

\usepackage[colorlinks, linkcolor = blue, citecolor = blue, urlcolor = blue]{hyperref}

\newcommand{\snl}{Sandia National Laboratories, Albuquerque, New Mexico 87185, USA}
\newcommand{\cint}{Center For Integrated Nanotechnology, Sandia National Laboratories, Albuquerque, New Mexico 87123, USA}
\newcommand{\unm}{Center for High Technology Materials and Department of Physics and
Astronomy, University of New Mexico, Albuquerque, New Mexico 87131, USA}

\begin{document}

\title{Fabrication of thin diamond membranes by Ne$^+$ implantation}

\author{Luca Basso}
\email{lbasso@sandia.gov}
\affiliation{\cint}

\author{Michael Titze}
\affiliation{\snl}

\author{Jacob Henshaw}
\affiliation{\cint}

\author{Pauli Kehayias}
\affiliation{\snl}

\author{Rong Cong}
\affiliation{Department of Physics, Brown University, Providence, Rhode Island 02912, USA}

\author{Maziar Saleh Ziabari}
\affiliation{\unm}

\author{Tzu-Ming Lu}
\affiliation{\cint}

\author{Michael P. Lilly}
\affiliation{\cint}

\author{Andrew M. Mounce}
\email{ammounce@sandia.gov}
\affiliation{\cint}

\begin{abstract}
Color centers in diamond are one of the most promising tools for quantum information science. Of particular interest is the use of single-crystal diamond membranes with nanoscale-thickness as hosts for color centers. Indeed, such structures guarantee a better integration with a variety of other quantum materials or devices, which can aid the development of diamond-based quantum technologies, from nanophotonics to quantum sensing. A common approach for membrane production is what is known as ``smart-cut'', a process where membranes are exfoliated from a diamond substrate after the creation of a thin sub-surface amorphous carbon layer by He$^+$ implantation. Due to the high ion fluence required, this process can be time-consuming. In this work, we demonstrated the production of thin diamond membranes by neon implantation of diamond substrates. With the target of obtaining membranes of $\sim$200 nm thickness and finding the critical damage threshold, we implanted different diamonds with 300 keV Ne$^+$ ions at different fluences. We characterized the structural properties of the implanted diamonds and the resulting membranes through SEM, Raman spectroscopy, and photoluminescence spectroscopy. We also found that a SRIM model based on a two-layer diamond/sp$^2$-carbon target better describes ion implantation, allowing us to estimate the diamond critical damage threshold for Ne$^+$ implantation. Compared to He$^+$ smart-cut, the use of a heavier ion like Ne$^+$ results in a ten-fold decrease in the ion fluence required to obtain diamond membranes and allows to obtain shallower smart-cuts, i.e. thinner membranes, at the same ion energy.
\end{abstract}

\maketitle

\section{Introduction}
\label{sec:intro}
Diamond is one of the leading material platforms for quantum information science (QIS) not only for its unique material properties, but for the presence of optically-active lattice defects, known as color centers. Indeed, these diamond defects' properties - such as spin-dependent optical transitions~\cite{doherty2013nitrogen,rogers2014all}, exceptional coherence times~\cite{balasubramanian2009ultralong,ohno2012engineering}, single photon emission~\cite{beveratos2002room,aharonovich2011diamond}, and controllable interaction with the surrounding environment~\cite{metsch2019initialization,rondin2014magnetometry, jiang2009repetitive} - have made them ideal candidates for different QIS applications like quantum computing~\cite{fuchs2011quantum, pezzagna2021quantum}, networking~\cite{ruf2021quantum}, and sensing~\cite{casola2018probing, schirhagl2014nitrogen, kehayias2022measurement,basso2022electric}. Most of these applications can benefit by the use of color centers embedded in a sub-micrometer thick, high-quality diamond membrane. For instance, these membranes can lead to an improved coupling of single emitters to waveguides or cavities~\cite{bhaskar2017quantum, faraon2011resonant, riedel2017deterministic}, an increased entanglement efficiency~\cite{pompili2021realization}, and a better interface with other materials for quantum sensing~\cite{yip2019measuring, pham2011magnetic}. In general, color-center-enriched diamond membranes can advance the development of diamond-based QIS technologies thanks to an improved integration with a variety of quantum materials or devices~\cite{guo2021tunable}. One of the most significant hurdles for the development of diamond-membrane-based quantum technologies is membrane production, since growth of high crystal quality, thin diamond film on non-diamond substrate is still challenging~\cite{gallheber2018growth, challier2018advanced}. For this reason, the most common technique for membrane production is diamond ``smart-cut'' used in combination with high-quality chemical vapor deposition (CVD) diamond overgrowth~\cite{guo2021tunable, lee2013fabrication, piracha2016Nanolett, aharonovich2012homoepitaxial, piracha2016scalable}. Smart-cut involves implanting a bulk diamond plate with He$^+$ ions to create an amorphous layer beneath diamond surface. Following this, a single-crystal, thin diamond film is grown over the implanted diamond. The amorphous layer is then selectively etched resulting in the exfoliation of the diamond membrane. Incorporation of color centers in the overgrown layer, either by doping during growth or subsequent ion implantation, has already been demonstrated for NV centers~\cite{hodges2012long} as well as other group IV defects such as SiV~\cite{pauls2020coupling}, GeV~\cite{guo2021tunable} and SnV~\cite{westerhausen2020controlled} centers with bulk-like spin coherence properties. However, due to the low mass of He, diamond smart-cut requires extremely high ion fluence, in the range $(0.5-1.5)\times 10^{17}$ ions/cm$^{2}$~\cite{guo2021tunable, gaathon2013planar}. 

To make this process more efficient, by reducing beam-time and cost, the smart-cut can be performed with a heavier ion, such as neon. In this work, we report the fabrication of ultrathin diamond membranes through Ne$^+$ implantation of diamond. We used Ne$^+$ instead of other heavier noble gases because smart-cut requires substrate atoms and irradiating ions to have comparable size. Larger ions, such as Ar$^+$, Kr$^+$ or Xe$^+$, do not penetrate the surface easily, leading to structural damage limited at the target surface resulting in an increased sputtering yield~\cite{evans1980some}. Moreover, Ne$^+$ does not introduce unwanted color centers in diamond, as Xe does~\cite{sandstrom2018optical}. To obtain $\sim$200 nm membranes we implanted diamonds with different ion beam fluences in the range $(1.38-11.0)\times 10^{15}$ ions/cm$^{2}$ with 300 keV energy. The implanted diamonds were characterized through SEM to observe the depth of the amorphous layer, and with Raman and photoluminescence (PL) spectroscopy to study the structural properties of the thin diamond layer capping the amorphous region. The diamond membranes were then isolated by removing the amorphous carbon layer through electrochemical (EC) etching in a boric acid solution. Finally, we estimated the damage threshold $D_c$, i.e. the minimum ion-induced vacancy density required to induce diamond amorphization, by addressing the discrepancies between the experimental results with Stopping Range of Ions in Matter (SRIM) simulations. We found that a target modeled as a two layers diamond/sp$^2$-carbon system better describes ion implantation with SRIM. From the obtained $D_c$ value, we estimated the lowest fluence required to obtain a smart-cut is $(2.3\pm0.1)\times 10^{15}$ ions/cm$^{2}$ (the lowest fluence smart-cut experimentally demonstrated in this work is of $2.75\times10^{15}$ ions/cm$^{2}$), more than an order of magnitude smaller compared to He$^+$ smart-cut. A faster and more efficient diamond membrane smart-cut production process could benefit and drive the development of a variety of diamond-based quantum technologies.
\begin{figure}[!b]
\centering
\includegraphics[width=0.9\linewidth]{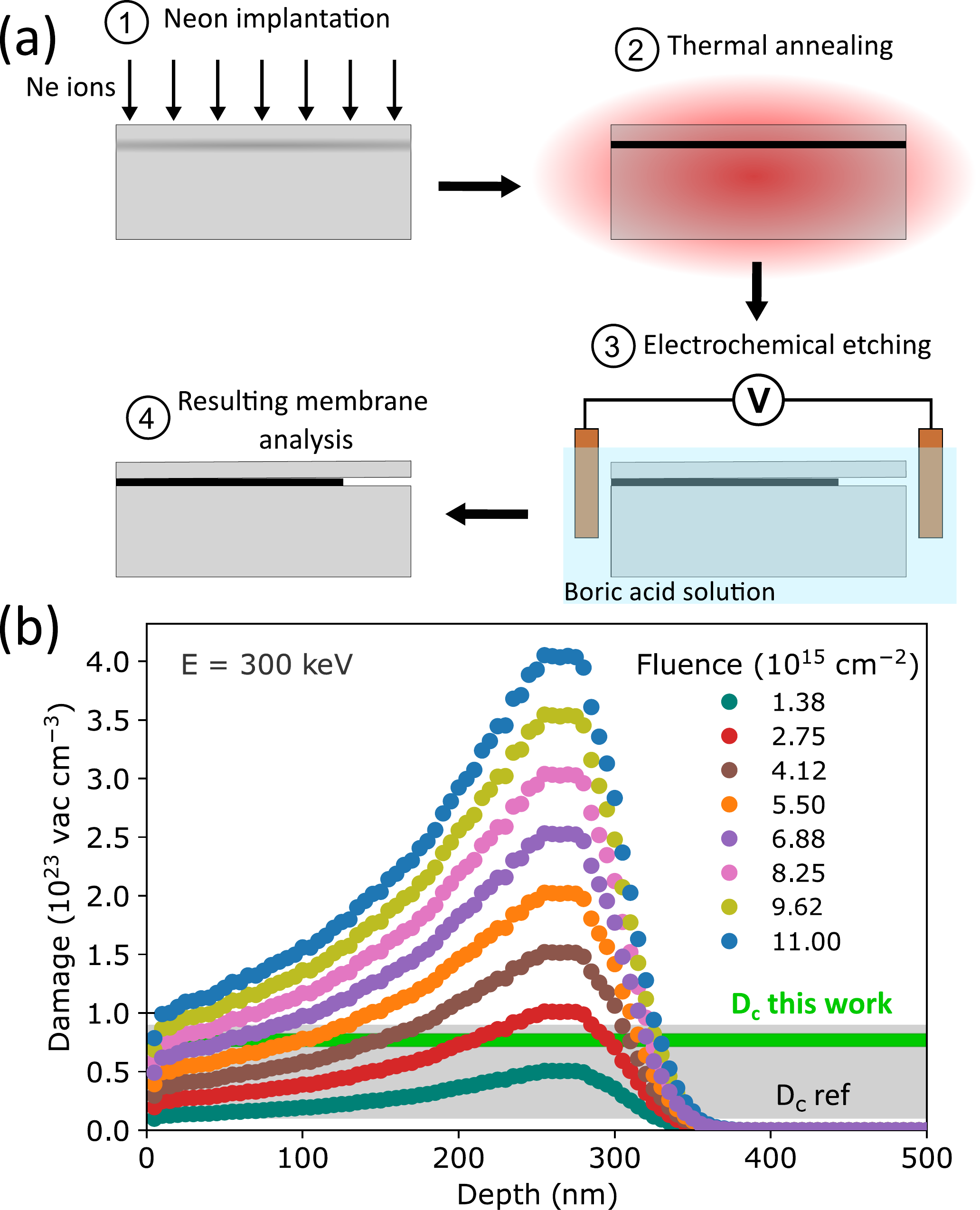}
\caption{(a) Schematics of Ne$^+$ smart-cut process. (1) Ne$^+$ implantation to form a damaged layer (dark grey layer) buried below diamond surface. (2) High temperature annealing to convert the damaged layer into an etchable amorphous sp$^2$ carbon phase (black layer). (3) Electrochemical etching in boric acid solution to etch the sp$^2$ carbon layer. (4) Membranes are partially exposed, i.e. the amorphous layer is removed only in a section of the sample, and still attached to the bulk diamond. (b) SRIM simulation of vacancy density profiles for a 300 keV Ne$^+$ implantation with different fluences. The critical damage threshold range $D_c = 1-9 \times 10^{22}$ vacancies/cm$^{3}$ for diamond He$^+$ implantation estimated from the literature is indicated by the grey box (${D_c}$ ref). The green line (${D_c}$ this work) is the value estimated in this paper of $D_c = (8.4 \pm 0.8) \times 10^{22}$ vacancies/cm$^{3}$.}
\label{Figure1}
\end{figure}
\begin{table*}[!ht]
  \begin{center}
    \begin{tabular}{c >{\centering}p{3cm} >{\centering}p{3cm} >{\centering}p{3cm} >{\centering\arraybackslash}p{3cm}} \toprule
      \textbf{Sample} & \textbf{Ne$^+$ fluence} & \textbf{Amorphous} & \textbf{Top diamond} & \textbf{Diamond}\\
       & \textbf{($10^{15}$ ions/cm$^{2}$)} & \textbf{layer (nm)} & \textbf{layer before EC etching (nm)} & \textbf{membrane (nm)}\\
      \midrule
      1 & 1.38 &  &  & \\
      2 & 2.75 & $262 \pm 12$ & $219 \pm 12$ & $222 \pm 11$\\
      3 & 4.12 & $338 \pm 12$ & $203 \pm 12$ & $195 \pm 10$\\
      4 & 5.50 & $363 \pm 12$ & $169 \pm 12$ & $176 \pm 11$\\
      5 & 6.88 & $613 \pm 15$ &  & $61 \pm 8$\\
      6 & 8.25 & $670 \pm 15$ &  & $55 \pm 6$\\
      7 & 9.62 & $676 \pm 15$ &  & $39 \pm 4$\\
      8 & 11.0 & $741 \pm 16$ &  & \\ \bottomrule
    \end{tabular}
    \caption{List of samples with the relative Ne$^+$ implantation fluence and details of SEM analysis before and after EC etching. For sample 1, the amorphous layer is not formed due to a low ion fluence. For samples 2-8 the diamond layer above the amorphous region, i.e. ``Diamond layer before EC etching'' could not be observed with SEM imaging. For sample 8 the amorphization reached the surface and no membrane is observed after the EC etching.}
    \label{tab:table1}
  \end{center}
\end{table*}
\section{Experimental Methods}
\subsection{Sample preparation}
\label{subsec:preparation}
\subsubsection{Ion implantation and thermal annealing}
Diamond membranes were obtained by implantation of 3$\times$3$\times$0.25 mm$^3$ optical-grade ($\leq 1$ ppm [N]) single-crystal diamonds produced by Element Six. Diamond plates were implanted with $^{20}$Ne$^+$ with an energy of 300 keV with a fluence ranging from 1.38$\times 10^{15}$ to 11$\times 10^{15}$ ions/cm$^{2}$ (see Table~\ref{tab:table1} for the list of all fluences). Implantation is performed using a 3 MV Pelletron accelerator using a plasma source containing a mixture of H and Ne. Samples are held perpendicular to the incident ion beam by means of carbon tape. The sample temperature is maintained $< 25^{\circ}$C through cooling by a liquid nitrogen cold-finger cryostat to avoid melting of the carbon tape from the high power of the ion beam. The ion beam is scanned over a 1/4" (6.35 mm) mask to ensure a homogeneous implantation profile. The fluence is measured through current integration off the copper sample holder. The samples were then annealed at 1200 $^\circ$C for 2h (heating rate of 2.5 $^\circ$C/min) in ultra-high vacuum.
\subsubsection{Electrochemical (EC) etching}
The implanted diamonds were glued in the center of a Petri dish, which was then filled with a saturated solution of boric acid in deionized (DI) water~\cite{fairchild2008fabrication}. Two copper electrodes with surface area $\sim$ 1 cm$^2$ were immersed in the solution and brought as close as possible to the two opposite sides of the diamond, to a distance between the anode and the cathode of $\lesssim$ 1 cm. We then used a DC power supply to apply 300 V between the electrodes. The concentration of the solution was adjusted by adding DI water or boric acid solution to maintain a current across the sample in the range $15-20$ mA. The samples were only partially etched, resulting in the membranes still being attached to the bulk diamond after the partial removal of the amorphous layer. After the etching, the diamonds were rinsed in acetone and then cleaned in a piranha solution to remove any contaminants from the EC etching.
\begin{figure*}[!ht]
\centering
\includegraphics[width=0.8\linewidth]{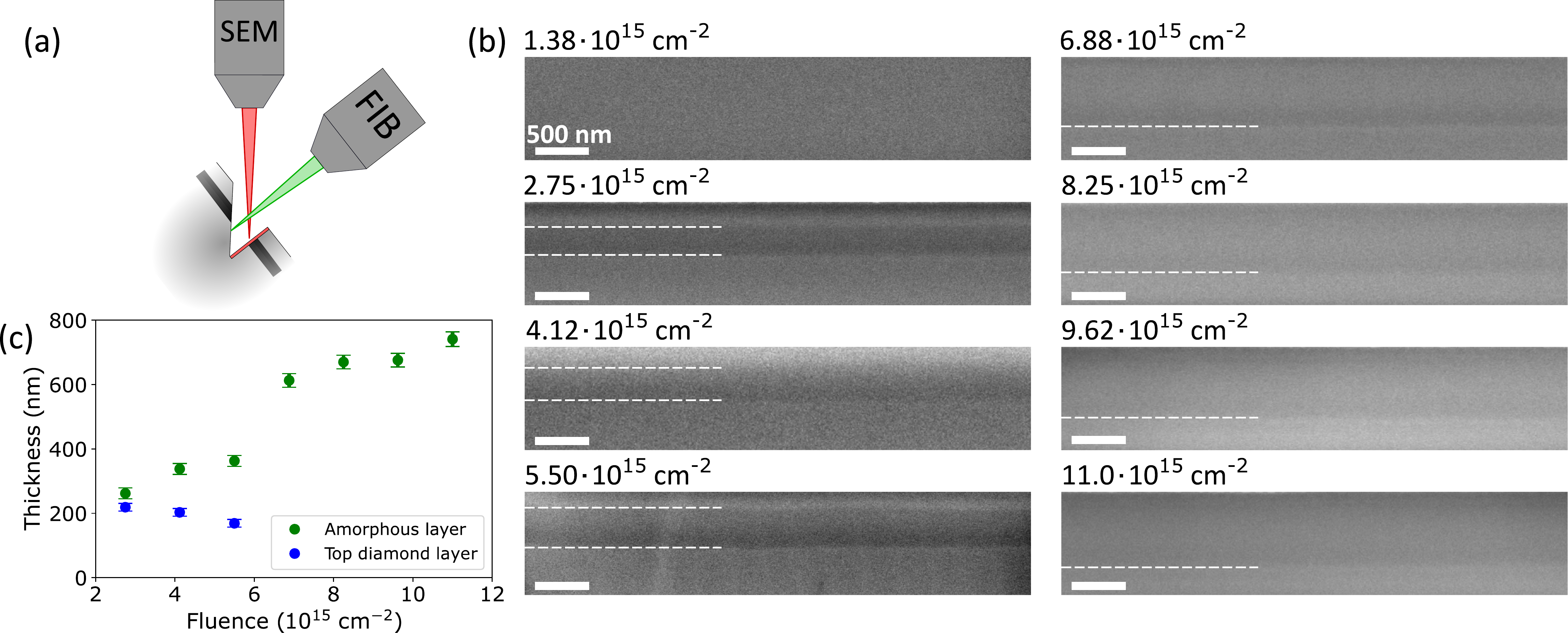}
\caption{(a) Schematics of the SEM imaging geometry. The samples are titled by a 52$^{\circ}$ angle, so that the surface is perpendicular respect to the FIB (green beam) that is used to mill the implanted diamonds and expose the amorphous layer (black layer). SEM (red beam) is then used to image the tilted samples so the cross section can be observed. (b) SEM images of the  smart-cuts obtained with different Ne$^+$ fluences (reported on top of each picture). White dashed lines indicate the amorphous layer position. Scale bar of 500 nm is the same for all the pictures. (c) Amorphous carbon layer thickness (green points) and thickness of the top diamond layer (blue points) measured from (b) as function of Ne$^+$ fluence. Values used in this plot are reported in Table~\ref{tab:table1}.}
\label{Figure2}
\end{figure*}
\subsubsection{SRIM simulation}
To simulate the defect density profile generated by Ne$^+$ implantation of diamond, we used Stopping Range of Ions in Matter (SRIM) Monte Carlo simulation software~\cite{ziegler2010srim}. To obtain the vacancy density profile we simulated 10,000 Ne$^+$ ions with 19.992 amu mass, 300 keV energy, and 500 nm target depth. The resulting vacancy density curve was then multiplied by the different ion fluences to obtain the different vacancy density profiles as function of Ne$^+$ implantation fluence. The diamond target was modeled as a $^{12}$C solid with: 3.51 g/cm$^{3}$ density, 37.5 eV atom displacement threshold energy, 3 eV lattice damage threshold, and 7.4 eV surface damage threshold~\cite{kehayias2021fitting}. To model the two-layer diamond/sp$^2$-carbon target with SRIM, we set the diamond with the same parameters as above, while the sp$^2$-carbon ones were changed into the standard SRIM parameters for graphite: 2.253 g/cm$^{3}$ density, 28 eV atom displacement threshold energy, 3 eV lattice damage threshold, and 7.41 eV surface damage threshold. 
\subsection{Sample characterization}
\label{subsec:characterization}
\subsubsection{SEM and FIB}
FIB milling and SEM imaging were performed with a FEI Nanolab 650 microscope. We used a Ga$^+$ FIB accelerated at 30 kV with a current of 3 nA to obtain a substantial diamond etching rate, whereas SEM was operated in secondary electron mode with an acceleration voltage of 5 kV. To reduce charging effects during the analysis, a layer of Pt was deposited over the region of interest before FIB milling. This layer also helped to clearly determine the position of the sample surface during SEM imaging of the diamond cross section, thus allowing to correctly determine the damaged layer depth.  
\subsubsection{Raman spectroscopy}
Raman and PL analysis were obtained with a WITec alpha300 Raman system. The excitation was provided by a 532 nm laser with 3.4 mW output power which was focused on the sample surface with a 100$\times$ 0.9 NA objective to a laser spot diameter of $\sim$ 20 $\upmu$m. The emitted light was dispersed with a 1800 g/mm grating and collected by a CCD camera. 
\section{Results and discussion}
\subsection{Ne smart-cut}
Schematics of the Ne$^+$ smart-cut process are given in Fig.~\ref{Figure1}(a). The diamond substrates were implanted with Ne$^+$ ions to form a damaged layer buried below diamond surface. We then annealed the samples to convert the region damaged above the critical damage threshold $D_c$ into an etchable sp$^2$ amorphous carbon layer. The substrates were finally electrochemically etched to selectively remove the amorphous carbon layer and release the diamond membrane. To obtain different membrane thicknesses, we smart-cut nine diamonds with different Ne$^+$ fluences at the same energy of 300 keV. We set the different fluences, ranging from 1.38$\times 10^{15}$ to 11$\times 10^{15}$ ions/cm$^{2}$, by comparing the SRIM simulation~\cite{ziegler2010srim} for Ne$^+$ implantation of diamond with the critical damage threshold range for He$^+$ implantation reported in the literature $D_c = (1-9)\times 10^{22}$ vacancies/cm$^{3}$~\cite{fairchild2012mechanism}, as shown in Fig.~\ref{Figure1}(b). The samples, labeled from 1 to 9, and the relative Ne$^+$ implantation fluence are reported in Table~\ref{tab:table1}. 
\begin{figure*}[ht!]
\centering
\includegraphics[width=0.8\linewidth]{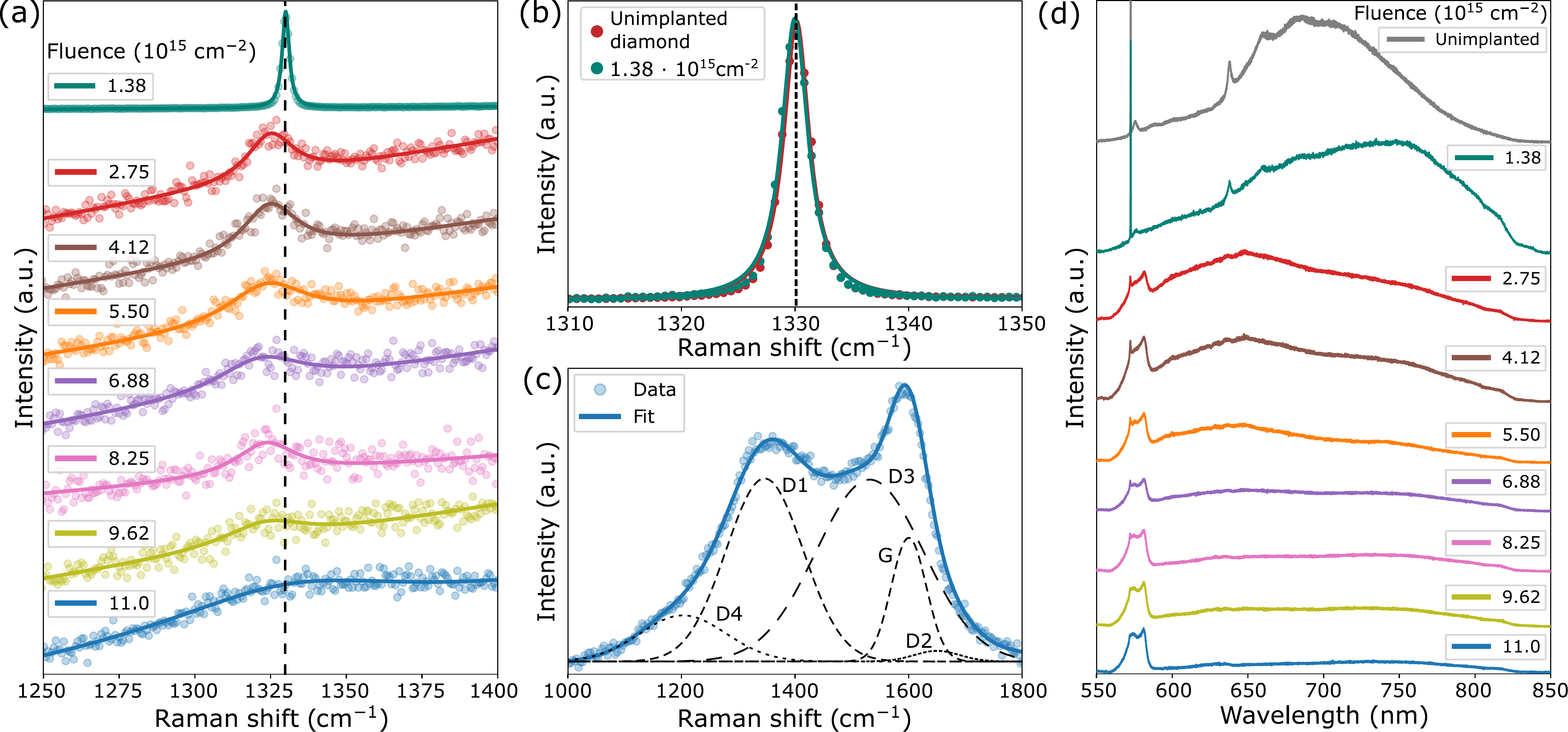}
\caption{(a) Raman spectra of implanted diamonds with different Ne$^+$ fluences. Solid lines are the Lorentzian fit of the experimental data (circles) with a linear background. Each curve is normalized respect to the maximum peak value. The black vertical dashed line is the position of the Raman peak for unimplanted diamond. (b) Comparison between the diamond Raman peak of sample 1 (dark green) and unimplanted diamond (dark red), both fit with a Lorentzian function. The Raman line for unimplanted diamond is centered at $1330.1 \pm 0.1$ cm$^{-1}$ with a FWHM of $2.7 \pm 0.1$ cm$^{-1}$, whereas sample 1 shows the Raman peak at $1329.9 \pm 0.1$ cm$^{-1}$ with a FWHM of $2.8 \pm 0.1$ cm$^{-1}$. (c) Raman spectrum of sample 9. Data points (light blue circles) are fit by a five peaks model~\cite{sadezky2005raman} with the peaks labelled as: G at $\sim$ 1600 cm$^{-1}$, D1 $\sim$ 1346 cm$^{-1}$, D2 $\sim$ 1648 cm$^{-1}$, D3 $\sim$ 1532 cm$^{-1}$, D4 $\sim$ 1202 cm$^{-1}$. (d) PL spectra from native NV centers in the Ne$^+$ implanted diamonds compared to the unimplanted one (grey curve). }
\label{Figure3}
\end{figure*}
\subsection{Implanted diamonds analysis}
\label{subsec:impl_diamond}
To measure the thickness and the depth of the amorphous layer, we FIB-milled the implanted substrate until the damaged layer was exposed, then used SEM to image the diamond cross section, as shown in Fig.~\ref{Figure2}(a). The SEM images of the implanted and annealed diamonds are reported in Fig.~\ref{Figure2}(b). Sample 1, implanted with the lowest fluence of 1.38$\times 10^{15}$ ions/cm$^{2}$, does not show the presence of a damaged layer, meaning that the damage threshold $D_c$ was not reached. In the other implanted samples the amorphous layers are visible as a darker layer~\cite{fairchild2012mechanism}, with thicknesses increasing with Ne$^+$ fluence from $262 \pm 12$ nm to $741 \pm 16$ nm, as shown in Fig.~\ref{Figure2}(c). For samples 2, 3 and 4, the top diamond layer thickness above the amorphous one ranges from $219 \pm 12$ nm to $169 \pm 12$ nm (see Table~\ref{tab:table1} for all measured values). For the remaining samples 5 to 8, the top diamond layer cannot be observed. This is likely a consequence of its very small thickness and the difficulty to SEM image small features in insulating materials. 
\begin{figure*}[ht!]
\centering
\includegraphics[width=0.8\linewidth]{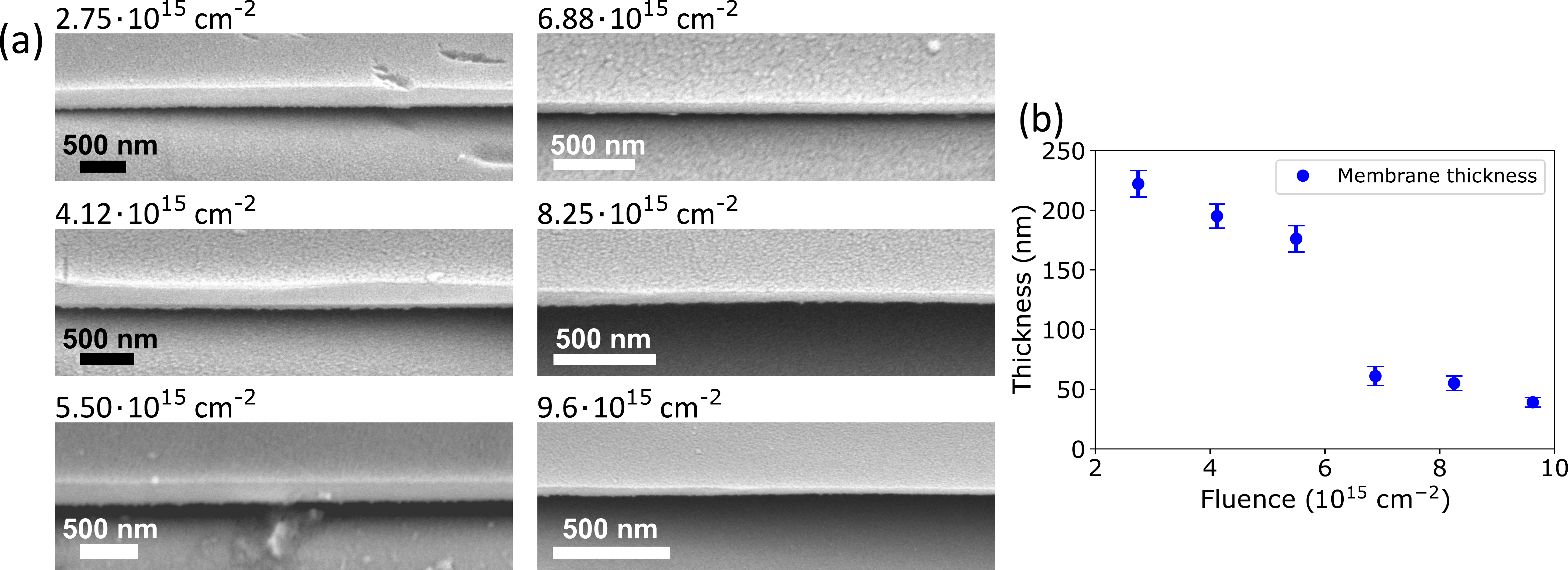}
\caption{(a) SEM images of the resulting diamond membranes obtained with different Ne$^+$ fluence (reported on top of each picture). (b) Membrane thickness (blue points) measured from (a) as function of Ne$^+$ fluence.}
\label{Figure4}
\end{figure*}
To determine if there is a diamond layer above the amorphous carbon layer, we used Raman spectroscopy and photoluminescence (PL) from the native NV centers. The Raman spectra are shown in Fig.~\ref{Figure3}(a): for each spectrum the fit (solid curves) of the experimental data is a Lorentzian peak with a linear background, while the vertical dashed black line is the peak position of unimplanted diamond at $1330.1 \pm 0.1$ cm$^{-1}$ (from Fig.~\ref{Figure3}(b)). While the sample 1 peak position is consistent with the peak position of unimplanted diamond, the other samples show a shift of the peak position towards smaller wavenumbers as a consequence of tensile stress acting on the diamond layer~\cite{di2013stress}. This stress may originate from the swelling of the implanted layer resulting from a density decrease after ion implantation~\cite{ orwa2000raman}. Indeed, diamond density reduction in proximity of the amorphous layer was previously observed in He$^+$ smart-cut diamond~\cite{fairchild2012mechanism} and, even if this decreased density is not enough to cause amorphization, it can cause lattice distortion and strain. The width of the peak also increases with Ne$^+$ fluence, as a result of increasing damage in the crystal structure~\cite{jamieson1995cross}. Moreover, with increasing fluence the peak intensity decreases, as the thickness of the diamond layer where the signal originates decreases. Samples 5, 6, and 7, for which a diamond layer above the amorphous one could not be observed with SEM, still show a diamond Raman peak (purple, pink, and light green data respectively), demonstrating the presence of a diamond layer at the sample surface. Sample 8 (light blue curve) does not show the presence of a peak that can be attributed to diamond phase. 
\begin{figure*}[!ht]
\centering
\includegraphics[width=0.8\linewidth]{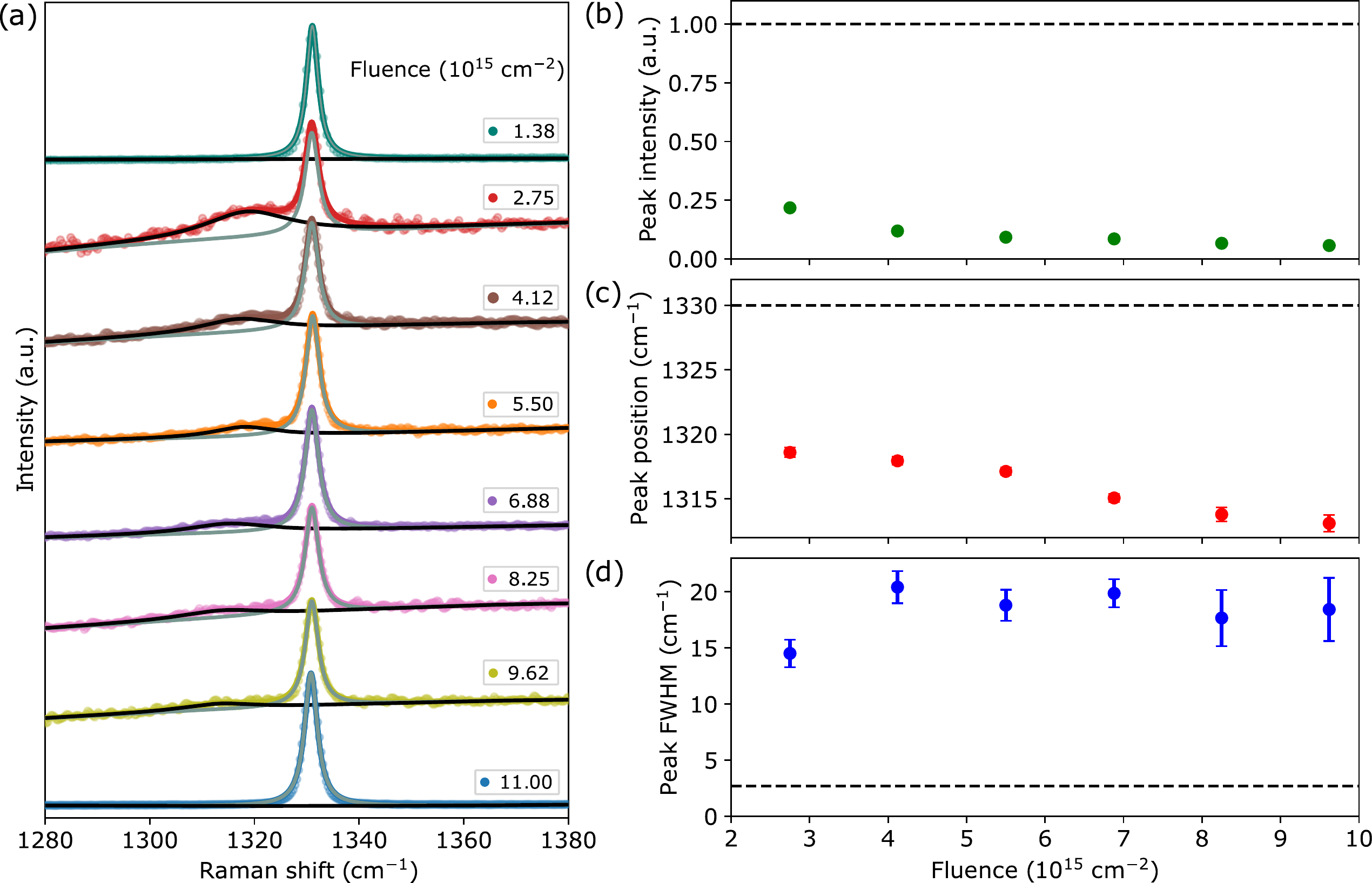}
\caption{(a) Raman spectra of the resulting diamond membranes. Experimental data (circles) are fit with two Lorentzian, one for the signal from the diamond substrate (grey curve at $\sim$ 1330 cm$^{-1}$) the other for the membrane Raman signal (black curve at $\sim$ 1315 cm$^{-1}$). The latter peaks parameters are reported as a function of Ne$^+$ fluence in (b) for peak intensity, (c) peak position, and (d) peak width. The three black dashed lines represent the values for unimplanted diamond with a Raman peak at $1330.1 \pm 0.1$ cm$^{-1}$ having FWHM of $2.7 \pm 0.1$ cm$^{-1}$. In picture (b) the peaks intensities for each sample are normalized respect to the peak intensity of unimplanted diamond. For (b-d) error bars are not visible when smaller compared to the data point size.}
\label{Figure5}
\end{figure*}
To confirm that there is no damage-induced phase transition occurring in sample 1, Fig.~\ref{Figure3}(b) shows its Raman spectrum (dark green curve) with an unimplanted diamond (dark red curve). The two peaks are completely overlapping, confirming the absence of damage and sp$^2$ carbon phase in sample 1. Fig.~\ref{Figure3}(c) reports the Raman spectrum on a wider range (1000-1800 cm$^{-1}$) for sample 8. The fit is a result of a five-Gaussian model, typical of amorphous carbon~\cite{sadezky2005raman}, where the five peaks (black curves labelled G, D1, D2, D3, and D4) originate from different sp$^2$ carbon atoms vibrational modes~\cite{ferrari2004raman}. The lack of a peak attributable to diamond in Fig.~\ref{Figure3}(c) demonstrates the absence of a diamond layer on top of the damage-induced amorphous region. 

Fig.~\ref{Figure3}(d) shows the PL spectra from the native NV centers of the implanted diamonds compared with the NV emission from an unimplanted diamond surface (grey curve). Sample 1 (dark green curve) still shows the zero phonon lines of NV$^0$ and NV$^-$ at respectively 575 nm and 637 nm, whereas the maximum of the characteristic phonon sideband, usually at $\sim$ 680 nm, is red-shifted to $\sim$ 750 nm. This can be attributed to the presence of neutral vacancy (GR1) defects in the implanted diamond~\cite{ozawa2018thermal, deak2014formation}. The sharp peak at 572.5 nm is the Raman signal of diamond (corresponding to 1330 cm$^{-1}$ at 532 nm laser excitation). The NV PL emission spectra from samples 2 to 8 are notably different. The PL emission from samples 2-4 is consistent with the NV$^0$ with phonon sideband maximum at $\sim$ 650 nm. The suppression of the NV$^-$ emission is consistent with the NVs being in proximity of sp$^2$ carbon atoms and is often observed from NVs in nanodiamonds~\cite{basso2019route, bradac2010observation}. The reason is the introduction of acceptor states by the sp$^2$ defects causing Fermi-level pinning below the NV$^-$/NV$^0$ transition~\cite{stacey2019evidence}, making the negative charge state of the NV unstable and the NV population dominated by NV$^0$. Furthermore, the PL intensity decreases with Ne$^+$ fluence as the active material, i.e. the diamond layer above the amorphous region, gets thinner, until a PL is detected but the NV features can no longer be identified, as observed for samples 5-9. Samples 2 to 9 also show the Raman features from the amorphous layer, the two broad peaks in the region $550 - 600$ nm ($615 - 2130$ cm$^{-1}$), superimposed over the sharp Raman peak of diamond at 572.5 nm. As already observed in the Raman spectra, the intensity of the diamond Raman peaks decreases with Ne$^+$ fluence. The absence of the Raman signal from amorphous carbon in the PL spectrum of sample 1 is an additional proof that the 1.38$\times 10^{15}$ ions/cm$^{2}$ implantation does not cause a phase transition and the increased vacancy concentration observed in the PL phonon sideband is not enough to damage the lattice.
\subsection{Diamond membrane characterization}
To release the diamond membrane, we removed the amorphous carbon layer by EC etching. The samples were partially etched, so that the membranes were still attached on the diamond substrates after the EC etching, as shown in Fig.~\ref{Figure1}(a). SEM images of the resulting membranes cross sections are shown in Fig.~\ref{Figure4}(a). The membranes thicknesses decrease with Ne$^+$ implantation fluence, from $222 \pm 11$ nm for sample 2, to $39$ $\pm$ $4$ nm for sample 8. Membrane thickness as function of Ne$^+$ fluence is reported in Fig.~\ref{Figure4}(b). Moreover, we could also observe that the membrane thickness measured before the EC etching, blue data in Fig.~\ref{Figure2}(c), matches the thickness of the final membranes for sample 2, 3 and 4. 

Results for the structural analysis performed with Raman spectroscopy are shown in Fig.~\ref{Figure5}(a). The Raman spectra are dominated by the signal coming from the diamond substrate, fit with a Lorentzian (grey curve) at $\sim$ 1330 cm$^{-1}$, as after the EC etching the membranes are laying on the bulk diamond substrate. The Raman signal originating from the membranes is the smaller peak at lower wavenumbers that is fit with a Lorentzian (black curve at $\sim$ 1315 cm$^{-1}$). Details of intensity, position, and width of these Raman peaks are plotted as function of Ne$^+$ fluence in Fig.~\ref{Figure5}(b), (c), and (d) respectively, together with the reference value of unimplanted diamond (dashed black lines) from Fig.~\ref{Figure3}(b). From Fig.~\ref{Figure5}(b) we could observe that the membrane Raman peak decreases with Ne$^+$ fluence, as expected due to a decreasing material thickness. Fig.~\ref{Figure5}(c) shows that the tensile stress is still present on the membranes, due to a shift of peak position towards smaller wavenumbers, as already observed in Fig.~\ref{Figure3}(a). We could observe that the shift increases with Ne$^+$ fluence, as for thinner membranes the strained section represent a larger portion of the sample, thus leading to an increasing shift in the Raman peak. Furthermore, the width of the membranes Raman peaks is $\sim10\times$ larger compared to the FWHM of the unimplanted diamond Raman peak, a signal of partially damaged crystal structures. Shifts and broadening of similar amplitudes have also been observed for Raman signal of He$^+$ smart-cut diamonds membranes~\cite{aharonovich2012homoepitaxial, magyar2011fabrication}. 
\begin{figure}[!ht]
\centering
\includegraphics[width=0.9\linewidth]{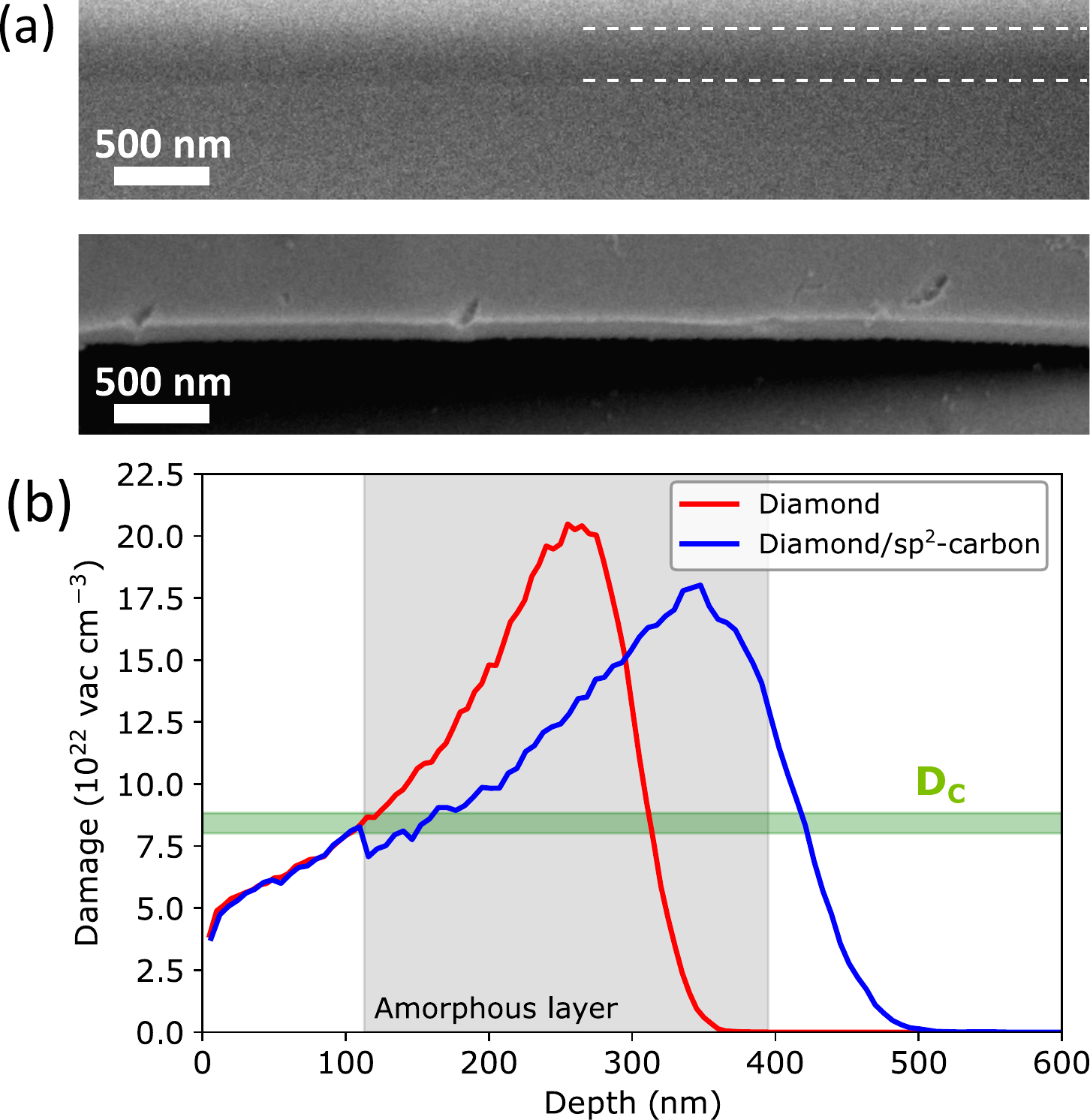}
\caption{(a) SEM images of diamond implanted with 300 keV angle at $5.50 \times 10^{15}$ ions/cm$^{2}$ with a 7$^{\circ}$ angle (top) and of the resulting membrane after EC etching (bottom). White dashed lines indicate the amorphous layer position. (b) Comparison between the measured amorphous layer thickness and depth (grey area) with two SRIM simulations: one for Ne$^+$ in diamond (red curve), one for Ne$^+$ implantation of a two layers target made of 113 nm of diamond and 500 nm graphite (blue curve). Green area is the estimated damage critical threshold of $D_c = (8.4 \pm 0.8) \times 10^{22}$ vacancies/cm$^{3}$.}
\label{Figure6}
\end{figure}
\subsection{Critical damage threshold estimation}
\label{subsec:critical}
When observing the amorphous carbon layer thickenss and depth, we could observe that simulations (Fig.~\ref{Figure1}(b)) and experimental results (Fig.~\ref{Figure2}(a)) do not match, in particular with SRIM underestimating the depth of the damaged layers. As a matter of fact, SRIM does not take into account some phenomena that may occur during ion implantation, such as channeling~\cite{jin2019channeling} or the change of material properties. To check whether channeling plays a relevant role, we implanted another diamond tilted by 7$^{\circ}$, with the same energy of 300 keV and a Ne$^+$ fluence of 5.5$\times 10^{15}$ ions/cm$^{2}$. The SEM cross-sectional images are reported in Fig.~\ref{Figure6}(a), after the implantation (top) and after the amorphous layer is etched (bottom). The amorphous layer is $282 \pm 13$ nm, capped by a diamond layer of $113 \pm 10$ nm; after the EC etching the thickness of the resulting membrane is $121 \pm 9$ nm. Fig.~\ref{Figure6}(b) compares this experimental result (grey area) with SRIM simulation of Ne$^+$ in diamond (red curve). Despite the 7$^{\circ}$ implantation angle, the simulation still underestimates the amorphous layer thickness, meaning that channeling does not play a major role. To further asses this discrepancy we performed a SRIM simulation, blue curve in Fig.~\ref{Figure6}(b), of Ne$^+$ implanted in a two-layer substrate made of 113 nm of diamond and 500 nm of sp$^2$-carbon with same energy and fluence. This two-layer target is a similar composition of the one expected at the end of the implantation process, and since sp$^2$-carbon has a lower density than diamond, this simulation should overestimate the damaged layer depth. This was actually observed, as shown in Fig.~\ref{Figure6}(b), which shows that the trailing edge of the amorphous layer lays in between the two SRIM simulation tails. Thus, SRIM inaccuracy in predicting the amorphous layer depth and thickness may be a consequence of the change in material physical properties and phase transition occurring during the implantation. Lastly, from comparing the SEM images and the SRIM simulations we could estimate the critical damage threshold $D_c$, resulting in $D_c = (8.4 \pm 0.8) \times 10^{22}$ vacancies/cm$^{3}$ (green area in Fig.~\ref{Figure6}(b)). This value is consistent with the literature reported results for $D_c$ in diamond, that for He$^+$ implantation is in the range $1-9 \times 10^{22}$ vacancies/cm$^{3}$~\cite{fairchild2012mechanism}. From this value of $D_c$ we used SRIM to estimate the minimum fluence $F_{min}$ for Ne$^+$ at 300 keV required to obtain a smart-cut (defined as the Ne$^+$ fluence resulting in a damage profile where the maximum value is equal to $D_c$), obtaining $F_{min} = (2.3\pm0.1)\times 10^{15}$ ions/cm$^{2}$.
\section{Conclusions}
In summary, we have demonstrated the fabrication of thin diamond membranes by Ne$^+$ implantation of diamond plates. The resulting membranes have tunable thicknesses, ranging from $39 \pm 4$ nm to $222 \pm 11$ nm, depending on the Ne$^+$ ions fluence. Moreover, by addressing the discrepancy between the experimental results and SRIM simulation, we obtained an estimate for diamond amorphization threshold. The use of a heavier ion, compared to the standard He$^+$ ``smart-cut'', has two main advantages. Firstly, it allows to reduce the implantation fluence required to form the amorphous layer by more then a factor 10. Furthermore, at the same ion energy Ne$^+$ has a smaller implantation depth compared to He, leading to a shallower damaged layer in turn resulting in a thinner diamond membrane. These two points, when considering that these smart-cut membranes are a sacrificial layer etched after being used as seeds for high-quality diamond overgrowth, show the relevance of our findings. A faster and more efficient method to produce thin diamond membranes to use as CVD templates for high-quality diamond overgrowth may benefit a wide range of diamond membranes applications, such as quantum information processing, nanophotonics, or diamond-based electronics.
\section*{Acknowledgements}
We thank Ania C. Bleszynski Jayich, Dirk Bouwmeester, Hyunseok Oh, and Ian Hedgepeth from UCSB for fruitful discussion. Sandia National Laboratories is a multi-mission laboratory managed and operated by National Technology and Engineering Solutions of Sandia, LLC, a wholly owned subsidiary of Honeywell International, Inc., for the DOE's National Nuclear Security Administration under contract DE-NA0003525. This work was funded, in part, by the Laboratory Directed Research and Development Program and performed, in part, at the Center for Integrated Nanotechnologies, an Office of Science User Facility operated for the U.S.~Department of Energy (DOE) Office of Science. This paper describes objective technical results and analysis. Any subjective views or opinions that might be expressed in the paper do not necessarily represent the views of the U.S. Department of Energy or the United States Government. 
\section*{References}
\bibliography{bibFile}
\bibliographystyle{iopart-num.bst}

\end{document}